\RequirePackage{ifpdf}
\documentclass{PoS}
\usepackage{epsfig}
\usepackage{upgreek}
\usepackage[authoryear,square]{natbib}
\bibpunct{(}{)}{;}{a}{}{,}

\title{Fast Transients at Cosmological Distances with the SKA}

\ShortTitle{Fast Radio Burst Science}

\author{\speaker{Jean-Pierre Macquart}$^1$,
Evan Keane$^2$,
Keith Grainge$^3$
Matthew McQuinn$^{4}$,
Robert Fender$^5$,
Jason Hessels$^6$,
Adam Deller$^{6}$,
Ramesh Bhat,$^{1}$,
Ren\'e Breton$^{7}$,
Shami Chatterjee$^8$,
Casey Law$^{4}$,
Duncan Lorimer$^{9}$,
Eran O. Ofek$^{10}$,
Malgorzata Pietka$^5$,
Laura Spitler$^{11}$,
Ben Stappers$^3$,
Cathryn Trott$^{1}$
\\

$^1$ICRAR/Curtin Institute of Radio Astronomy, Bentley, WA 6845, Australia\\
        E-mail: \email{J.Macquart@curtin.edu.au} \\
$^2$Swinburne University of Technology \\
$^3$University of Manchester \\
$^4$UC Berkeley \\
$^5$University of Oxford \\
$^6$ASTRON \\
$^{7}$University of Southampton \\
$^8$Cornell University \\
$^{9}$West Virginia University \\
$^{10}$Weizmann Institute of Science \\
$^{11}$MPIfR \\
}

\abstract{

Impulsive radio bursts that are detectable across cosmological distances constitute extremely powerful probes of the ionized Inter-Galactic Medium (IGM), intergalactic magnetic fields, and the properties of space-time itself.    Their dispersion measures (DMs) will enable us to detect the ``missing'' baryons in the low-redshift Universe and make the first measurements of the mean galaxy halo profile, a key parameter in models of galaxy formation and feedback.  Impulsive bursts can be used as cosmic rulers at redshifts exceeding 2, and constrain the dark energy equation-of-state parameter, $w(z)$ at redshifts beyond those readily accessible by Type Ia SNe.  Both of these goals are realisable with a sample of $\sim 10^4$ fast radio bursts (FRBs) whose positions are localized to within one arcsecond, sufficient to obtain host galaxy redshifts via optical follow-up.  It is also hypothesised that gravitational wave events may emit coherent emission at frequencies probed by SKA1-LOW, and the localization of such events at cosmological distances would enable their use as cosmological standard sirens. 

To perform this science, such bursts must be localized to their specific host galaxies so that their redshifts may be obtained and compared against their dispersion measures, rotation measures, and scattering properties.  The SKA can achieve this with a design that has a wide field-of-view, a substantial fraction of its collecting area in a compact configuration (80\% within a 3\,km radius), and a capacity to attach high-time-resolution instrumentation to its signal path.
}

\FullConference{
Advancing Astrophysics with the Square Kilometre Array\\
June 8-13, 2014\\
Giardini Naxos, Italy}

\begin{document}

\section{Observational Phenomenology of Fast Radio Bursts}

Before discussing the cosmological science that can be achieved by
observing fast radio bursts (FRBs), it is important to describe them as an
observational phenomenon.  Radio pulsar surveys sample the sky at high time
resolution (typically 100-$\upmu$s samples) and search in dispersion
measure (DM) space.  The dedispersed time-frequency data are searched for
both periodic signals as well as individual impulsive events, i.e. single
pulses, with durations from the sampling time up to at least tens of
milliseconds.  The single-pulse searches are sensitive to bright pulsars
and those that emit only sporadically, such as the sub-class known as the
rotating radio transients \citep[RRATs;][]{McLaughlin06}.  Individual
pulses can be differentiated from radio frequency interference (RFI)
because they are localized on the sky (appearing in only one beam of a
multi-beam system, or imaged using fast-imaging techniques); broadband; and
have DMs that indicate that the signal has been dispersed by an intervening
ionized material with a delay proportional to $\nu^{-2}$ (cold plasma
dispersion).  RFI almost never exhibits similar dispersive properties.
When combined with a model for the free electron distribution of the
Galaxy, the DM can also be used to estimate the distance to the emitting
source \citep[e.g.][the NE2001 model]{CL01}.  

Recent pulsar surveys have discovered close to a hundred RRATs\footnote{see http://astro.phys.wvu.edu/rratalog/}, almost all of which have shown multiple detected pulses, and have DMs that are less than the maximum Galactic DM expected from the NE2001 model along that line-of-sight.  It appears likely that all the RRATs are sporadically-emitting neutron stars.  Interestingly, such searches have also discovered a small number of individual pulses that are now commonly referred to as FRBs.  What distinguishes FRB signals from those of pulsars/RRATs is that they are apparently non-repeating and have DMs that are well in excess of the maximum Galactic contribution predicted by NE2001.  The high DMs in particular have led to the conclusion that these signals are likely of extragalactic origin and the dispersive delay results from the combined interstellar medium of our Galaxy, the intergalactic medium and that of a potential host galaxy.  The apparently non-repeating nature of the signal also suggest some kind of cataclysmic event, though there is no way to conclusively show that the signal will never repeat.

Determining the sources of the FRB signals, and definitely proving that they are of extragalactic origin has been hampered by their relatively poor sky localization.  The positional uncertainty of the known FRBs is on the order of arcminutes, while arcsecond localization is needed to associate them with a known optical source or host galaxy.  Such localization requires at least several kilometer baselines, which are not currently used in ongoing pulsar surveys.

In this paper we discuss the utility of FRBs --- the events originally termed ``Lorimer bursts'' --- and other as yet undiscovered impulsive extragalactic phenomena as cosmological probes.  Our discussion is limited purely to their cosmological applications, and does not address the mystery of their origin.   Although understanding the origin of FRBs is highly compelling, the overwhelming abundance of suggested explanations makes it difficult to specify the astrophysical implications of their identification.  
The layout of the remainder of this paper is as follows.  In \S \ref{sec:cosmo} we describe the various facets of cosmology that are potentially addressed using observations of FRBs.    In \S \ref{sec:require} we describe the requirements driven by the science goals motivated in the previous section, while in \S \ref{sec:SKA1}, \S \ref{sec:reducedSKA} and \S \ref{sec:fullSKA}, we describe respectively the portions of the science case that can be addressed with SKA1, a reduced-sensitivity SKA1 and, finally, the full SKA.  Our conclusions are presented in \S \ref{sec:conclusions}.

\section{Extragalactic Transients as cosmological probes} \label{sec:cosmo}

Two key questions facing early twenty-first century cosmology are the nature of the mysterious dark energy that appears to pervade 70\% of the mass-energy of the Universe, and the distribution of the large reservoir of missing baryonic matter in the present Universe.  The detection of impulsive bursts at cosmological distances offers an entirely new means of solving both of these problems.  

Over the past seven years the realization has grown that cosmological bursts may be common and bright.  The class of events known as Fast Radio Bursts (FRBs), discovered with the Parkes and Arecibo multibeam receivers \citep{Lor07, Thorn13, Spitler14}, represents the first of a new category of radio transients detected at apparently cosmological distances.  These events are bright ($>1\,$Jy\,ms), and their millisecond durations make it possible to directly measure the column density of ionized plasma in intergalactic space via their frequency-dependent time of arrival (i.e. their dispersion measure).   The dispersion measures of these bursts place them {\it prima facie} at redshifts out to $>1$.  

The short durations of these bursts makes them exquisite probes of the dispersion measure, enabling us to account for every single ionized baryon that lies between the burst and the Earth to high precision, and to measure the curvature of spacetime through which the radiation propagates on its way to Earth.
This unique attribute has led to a spate of recent suggestions \citep{ MK13, Deng14,Zhou14, McQ14} describing their utility as cosmological probes.  These fall into several categories:
\begin{itemize}
\item Locators of the ``missing'' baryons in the low ($z \leq 2$) redshift universe \citep{Ioka03, Inoue04,Deng14,McQ14};
\item High-redshift cosmic rulers which have the potential to determine the equation-of-state parameter $w$ over a large fraction of the history of the Universe \citep{Zhou14} and
\item Potential probes of primordial (intergalactic) magnetic fields and turbulence \citep{MK13}.
\end{itemize}
In this paper we discuss each of these science cases below in detail.  For the purposes of illustration we shall primarily frame the discussion in terms of FRBs, but we stress that such investigations would be possible using any class of impulsive extragalactic transients.  With the modest sensitivity and narrow field of view of the telescopes with which they were detected, FRBs likely represent only a tiny fraction of all the classes of new transient phenomena awaiting discovery in the SKA-era.  

\subsection{Missing Baryons}
The ``missing'' baryon problem refers to the fact that most baryons in the low-redshift Universe remain undetected \citep{CO99, CO07, Bregmann, Shull, McQ14, Sullivan}.  Only 5\% of the cosmic baryons at $z \sim 0$ lie within galaxies, another 5\% are detected in X-ray coronae in massive groups and clusters, and 30\% reside in a warm intergalactic phase observed in Ly$\alpha$ absorption.  The constraints on the remaining 60\% of the baryons are weaker, and they are believed to reside at temperatures and densities that are difficult to detect via traditional spectral line absorption and emission diagnostics \citep[however see \cite{Gnat}]{CO99}.

Most dark matter is found within massive galaxy halos, but most baryonic matter is outside this scale ($\gtrsim 100\,$kpc). Since baryonic matter is more sensitive than dark matter to star formation and AGN feedback in galaxies, the baryon distribution in cosmological simulations is very sensitive to the feedback implementation.  The location of these missing baryons constitutes an important element of galaxy halo accretion and feedback models.  Although half of the Universe's dark matter lies within halo masses $>10^{10} {\rm M}_\odot$, less than half of the baryonic matter lies within these halos.  The unseen baryons associated with halos below $\sim 10^{13}\,{\rm M}_\odot$ likely must lie outside of the halo virial radius because a hot atmosphere containing these baryons would be thermally unstable, which would lead to a cooling flow that would cause an (unobserved) excess of star formation in these galaxies.

The dispersion measures of FRBs can resolve the origin of these missing baryons.  The distribution of dispersion measures is sensitive to the locations of these baryons, and can determine whether they lie within the virial radius of $10^{11}$-$10^{13}\,{\rm M}_\odot$ halos, or whether they lie further out in an intra-halo medium.  

The basis of this determination is the variation in dispersion measure for a set of FRBs at similar redshifts.  The sightline-to-sightline scattering in dispersion measure observed is predicted to be primarily caused by the scatter in the number of collapsed systems encountered along the ray path to an FRB.  The measured DM of an FRB contains four contributions, 
\begin{eqnarray}
{\rm DM}_{\rm tot} = {\rm DM}_{\rm MW} + {\rm DM}_{\rm IGM} + {\rm DM}_{\rm host} + {\rm DM}_{\rm FRB}, 
\end{eqnarray}
namely from the Milky Way, the IGM, the host galaxy and the circum-burst shell (if any) respectively.  The Milky Way contribution is often known from pulsar models and is small (typically $\sim 50\,$pc\,cm$^{-3}$ off the Galactic plane), and the host contribution is likely similarly small for all bursts not located near the centres of their host galaxies.  Moreover, both the host and FRB contributions\footnote{The FRB DM contribution is limited by the fact that it cannot be so dense as to prevent propagation of the radiation (i.e. the plasma frequency must not exceed the radiation frequency), which limits the emission to regions with densities $\lesssim 10^{10}\,$cm$^{-3}$.} to the DM decrease as $1/(1+z)$ relative to local contributions,  where $z$ is the FRB host galaxy redshift.

\begin{figure}[h!]
\centerline{\epsfig{file=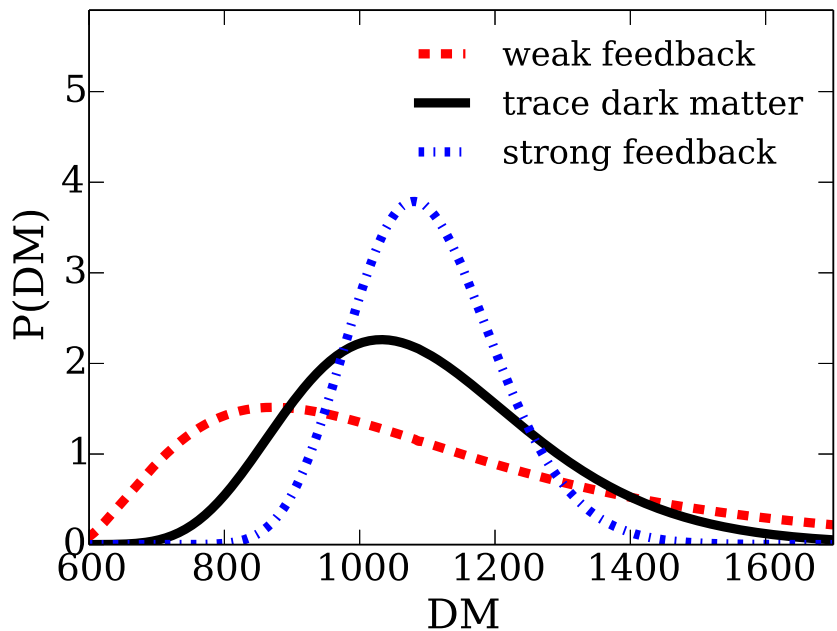, scale=0.7}}
\caption{Possible probability distributions of FRB dispersion measures for bursts located at $z=1$.  The distribution depends on how the baryons are distributed near the halos of galaxy clusters along the line of sight.   The more diffuse the gas, the more concentrated is the probability density around its central value.  Here, strong feedback corresponds to a scenario in which the baryonic extent of each galaxy cluster halo extends to 2 virial radii, while the weak feedback corresponds to one in which the halo extends to only half the cluster virial radius.  (See \citet{McQ14} for more details.)
} \label{DMpdfsFig}
\end{figure}

A key observable quantity is the DM probability distribution function.   For a set of FRBs at the same redshift, the shape of the DM distribution function depends sensitively on the extent of the distribution of the baryons around the halos.  Figure \ref{DMpdfsFig} shows how the DM distribution function of a sample of 1000 FRBs at $z=1$ depends on the nature of the distribution of baryons around their halos.
Note that all models predict a high-DM tail in the distribution.  A general property is that the distribution function is more centrally peaked if the gas around the halo is more diffuse or if the halos capable of retaining their gas are rarer.  

Another method of exploring the baryon distribution involves stacking FRBs based on their angular proximity to galaxies in order to measure the mean baryon profile of galaxies to large radii (McQuinn 2014).  This technique requires sub-arcminute (or better) localization of each FRB.

\subsection{Cosmic Rulers}
Measurements of Type Ia SNe out to $z \sim 1.5$ have been used to determine the dark energy content of the Universe.  The opportunity exists to make much more detailed measurements of the geometry of the Universe using impulsive transients at redshifts $>2$, where their DM contribution is dominated by the IGM.  FRBs are much brighter relative to telescope sensitivity than Type Ia SNe and are potentially easily detectable to much higher redshifts, especially with high-sensitivity, wide-field telescopes.  FRBs offer access to the dark energy equation of state parameter $w(z)=p/\rho$ \citep{Zhou14}.  

The basis for using transients as cosmic rulers is that its average ${\rm DM}$ at a redshift $z$ depends on the geometry of the Universe in a specific manner.  Transients are usable as cosmic rulers in the sense that a comparison of their DM to their redshift, when coupled with a model for the average electron density of the Universe as a function of redshift, enables one to measure the time of flight of the photons and hence the {\it path length}.  It is in this sense that transients represent cosmic rulers\footnote{The essence of the argument is that, in the integral over path length $\int n_e dl$, one writes $dl = c dt = c |dt/dz| dz$.}.   The exact relation between DM and $z$ is \citep{Zhou14}:
\begin{eqnarray}
\langle {\rm DM}_{\rm IGM}(z) \rangle = \Omega_b \frac{3 H_0 c}{8 \pi G m_p} \int_0^z \frac{(1+z') f_{\rm IGM} \left[ \frac{3}{4} X_{e,H}(z') + \frac{1}{8} X_{e,He}(z') \right] }{\left[ \Omega_{\rm M} (1+z')^3 + \Omega_{\rm DE}(1+z')^{3[1+w(z')]}  \right]^{1/2}} dz'  \label{DMzeq}
\end{eqnarray}
where  $\Omega_b$, $\Omega_{\rm M}$ and $\Omega_{\rm DE}$ are the baryonic, matter and dark energy densities, respectively, relative to the critical density, $\rho_c = 3 c^2 H_0^2/8 \pi G$.  $X_{e,H}$ and $X_{e,He}$ are, respectively, the ionization fractions of Hydrogen and Helium.  

\begin{figure}[htbp!]
\begin{center}
\epsfig{file=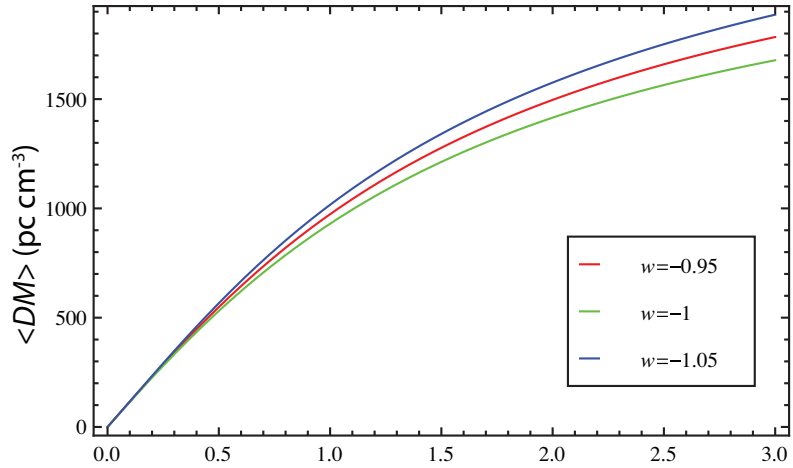,scale=0.8}
\caption{The mean DM contribution to the IGM of an FRB as a function of redshift based on eq.(2.2) for a concordance $\Lambda$CDM Universe with $\Omega_{\rm M}=0.318$, $\Omega_\Lambda=0.682$, $\Omega_b =0.049$ and $H_0 = 70$\,km\,s$^{-1}$Mpc$^{-1}$. } \label{DMrelationFig}
\end{center} 
\end{figure}

The determination of cosmological parameters relies upon the detection of a sufficient number of FRBs that it is possible to measure the {\it average} DM of FRBs as a function of redshift.  Figure \ref{DMrelationFig} shows the expected scaling of DM with redshift for a concordance cosmology.  The main contaminant is the uncertain contribution ${\rm DM}_{\rm host}$ and ${\rm DM}_{\rm FRB}$ to the total dispersion measure.  However, these contributions diminish relative to the local contribution as $1/(1+z)$ and rapidly decrease relative to an IGM whose mean density increases as $(1+z)^3$.  This makes the technique viable for FRBs at $z \gtrsim 2$.  Zhou et al. (2014) estimate that $\sim 10^3$ FRBs must be detected in order to place significant cosmological constraints on $w$, as demonstrated in Figure \ref{DMrelationFig}.

\subsection{Intergalactic Magnetic Fields and Turbulence}

Two of the seven FRBs reported in the literature so far exhibit clear evidence for temporal smearing caused by scattering. The observed few-millisecond timescale of the smearing however, cannot be accounted for by the interstellar medium of the Milky Way, whose maximum expected contribution is at most microseconds for the lines of sight through which these bursts were detected.  The origin of this scattering therefore lies in the turbulent intergalactic medium, or in the turbulent interstellar medium of the burst host galaxy.  The IGM and host-galaxy scattering scenarios can be distinguished on the basis of the redshift dependence of the magnitude of the scattering \citep{MK13}.

If the scattering is inherent to the IGM, the effect affords a means of probing the evolution of energy deposition in the IGM.  The sources invoked to explain reionization of the Universe and feedback associated with galaxy formation inject energy into the IGM on large scales, driving a turbulent cascade that should drive the evolution of inhomogeneity in the IGM.  Sources of energy input include the radiative and mechanical energy of AGN and their jets, the UV and X-ray emission and stellar winds from young stars and the flows driven by supernovae \citep[and references therein]{CO99, CO07}.

The combination of DM and scattering information affords the possibility of tomographically reconstructing the structure and turbulent properties of the Universe's baryonic content, similar to the manner in which pulsar measurements have been used to map out the structure of the Milky Way's interstellar medium \citep[e.g.][]{ARS95, TC93, CL01}. 

However, there are arguments that the observed scattering may not be inherent to the IGM unless it is highly clumped in relatively dense regions, or the physics of the turbulent cascade is far from that envisaged in simple models \citep{MK13, LG14}.  In this case, the temporal smearing must originate in the ISM of the host galaxy or in intervening dense, gravitationally collapsed systems, and the smearing provides a rare opportunity to probe turbulence in external galaxies.  

Most coherent mechanisms that generate extremely bright emission of the sort observed in FRBs generate radiation that is partially polarized (e.g. pulsar emission).  If FRB emission contains a significantly linearly polarized component, it affords a means of probing intergalactic magnetic fields.  Comparison of the DM and rotation measure \citep[RM; see e.g.][]{Kulkarni14} of the transient will enable measurements of the mean magnetic field of the IGM, in much the same way that RM measurements of pulsars enable measurements of the in situ magnetic field of our own Galaxy \citep{Noutsosetal,Han}.

\subsection{The future of Fast Transients Research in the SKA era}

\begin{figure}[h!]
\centerline{\epsfig{file=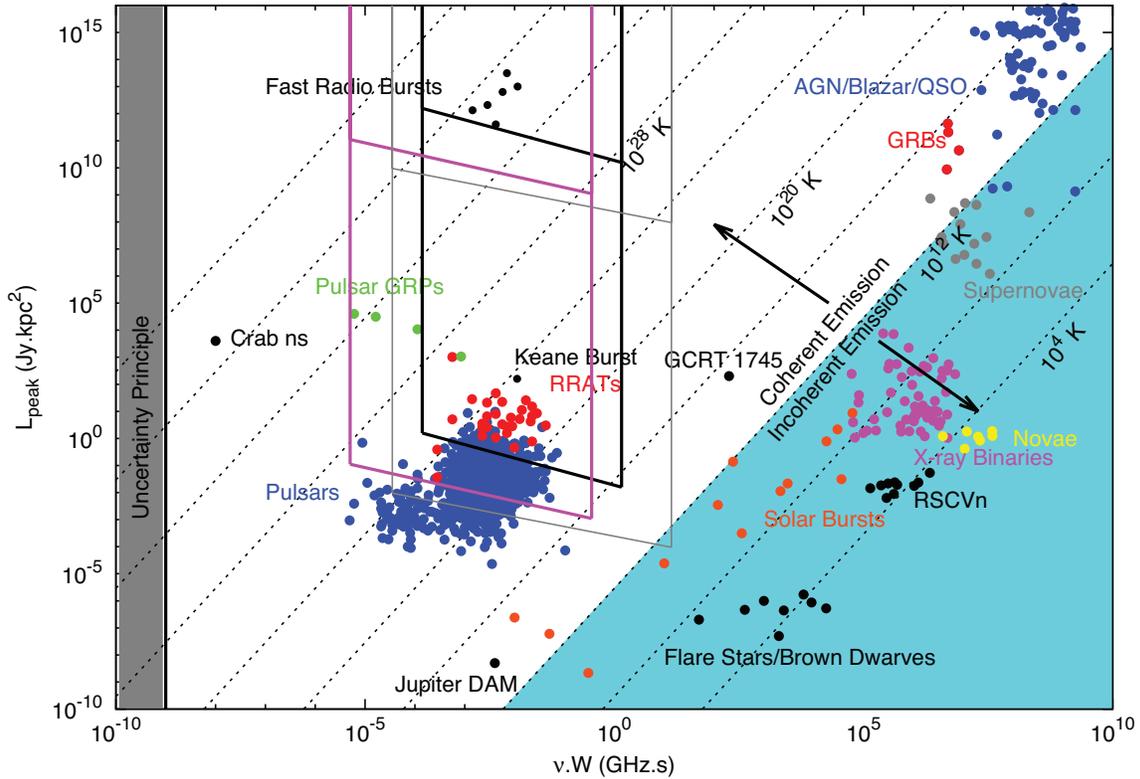,scale=0.6,angle=270}}
\caption{Transient parameter space of specific luminosity versus the
product of observing frequency and transient duration. All the known
sources of transient radio emission are shown. Overplotted are $1$~kpc
and $1$~Gpc sensitivity curves for Parkes (black), SKA1-LOW (pink) and
SKA1-MID (grey). This figure illustrates the depth to which SKA1
can probe the transient radio sky. For example the published FRBs are
all just above the Parkes $1$~Gpc sensitivity limit (black
line). These sensitivity limit curves are for beamformed
searches. Fast imaging transient searches extend the RHS boundary of
these curves all the way to the far right extent of the plot.  This Figure is based on a representation of parameter space originally made by \citet{cordes04}.
} \label{fig:transient_fig}
\end{figure}

Figure~\ref{fig:transient_fig} shows the transient parameter space
often used to characterise the known sources in the transient radio
sky. One can see that the known sources span $25$ orders of magnitude
in radio (pseudo-)luminosity and $18$ orders of magnitude in terms of
$\nu*W$, covering timescales ranging from weeks, for AGN, to nanoseconds, which is the Nyquist rate for giant pulses from the Crab pulsar. Despite this, the parameter space is mostly devoid of sources.  While past discoveries are no guarantee of future successes, it is nonetheless remarkable that whenever we have probed a new area in this
parameter space we have found new sources, each bringing their own
astrophysical applications, as well as the huge potential for
discovery that the SKA has within the key projects of the Transients
SWG.  Several design aspects of the SKA render it an exceptionally potent transients exploration machine: aside from its high sensitivity, and thus ability to probe to large distances, it will have a hugely increased likelihood of detection by virtue of its large field-of-view and observing-time product. This is a result of the multi-beaming capabilities of SKA-LOW and SKA-MID, and the fact that transient
detection systems will operate in a `piggyback' mode on all SKA
projects 24/7/365. This alone increases the equal-sensitivity rate of
detection, by many orders of magnitude, and will ensure that the SKA
will operate as an unprecedented all-sky all-timescale monitor.

\section{Science-driven Telescope Requirements} \label{sec:require}
%
%
%
%
%
%
%

The basic requirements resulting from the science cases mentioned above are as follows:-
\begin{itemize}
\item For the missing baryon science we need to detect of order $10^2$ events per redshift bin.  This translates to requiring the detection of order $10^4$ events in total (depending on their redshift distribution.)
\item It is estimated that the cosmic ruler science requires at least $10^3$ events.  The exact number greatly depends on the dispersion of FRB DM values about the mean as a function of redshift.  It also depends on the contribution of the host galaxy to the overall DM which is, as yet, undetermined.
\item The science depends crucially on our ability to determine FRB redshifts, requiring event localization to within $\sim$0.1-0.5$^{\prime \prime}$ for events at $z \gtrsim 1$.  The Dark Energy Survey (DES) and surveys with LSST will enable us to obtain redshifts for these counterparts\footnote{Experience at sub-mm wavelengths suggest that 0.1$^{\prime \prime}$ localization is necessary to achieve unambiguous galaxy identifications.  The point spread functions of DES and LSST have sizes of 0.9$^{\prime \prime}$ and 0.7$^{\prime \prime}$ respectively; localization to $0.1^{\prime \prime}$ is possible with sufficient S/N as long as confusion is not an issue.}.  We elaborate on this point in \S\ref{sec:localize}.
\end{itemize}
A crucial point is that a large amount of time on sky is necessary to achieve the large numbers of event detections needed to prosecute the cosmological investigations outlined above.   This argues strongly that surveys for FRBs need to operate commensally with other telescope projects.  One could expect at most 10\% of the telescope time to be allocated to a dedicated survey for FRBs, whereas a survey capable of operating in conjunction with other projects could feasibly achieve close to 100\% efficiency, and could build up a sufficiently large FRB sample much more rapidly.

To put this remark in perspective, we note the estimated event rate based on Parkes detections is $10_{-4}^{+6} \times 10^3$~events\,sky$^{-1}$day$^{-1}$ (at a centre frequency of 1340\,MHz and with an effective 1\,ms timescale 10$\sigma$ detection threshold of 0.46\,Jy).  Parkes therefore detects a single FRB on average once every 10\,days of telescope time.  Even if this telescope were dedicated solely to FRB surveys, the detection rate is three orders of magnitude too small to successfully the pursue cosmological studies with FRBs.  Though the expected event detection rate on SKA1 is higher (e.g. we see below that SKA1-SUR's detection rate is $\sim 200$ times higher than that of Parkes), only a fully commensal survey is capable of detecting FRBs in sufficient numbers.

\section{Science with SKA1} \label{sec:SKA1}
Here we outline several ways in which the detection {\it and follow-up} of cosmological events may be prosecuted with SKA1.  At present a great deal of important information, such as FRB spectral indices and source counts, are unknown.  The effect of scatter broadening due to turbulence in intervening plasmas, which scales as $\nu^{-4}$ if it is relevant, is the biggest unknown in determining the optimum survey frequency for cosmological transients.  As a result, it is difficult to design a survey strategy for FRBs -- or any other type of yet-undiscovered cosmological transients -- so we advocate a flexible approach and investigate several different means of pursuing this science.  

\subsection{SKA1-LOW}

The primary advantages of using SKA1-LOW for fast transients detection are its large field of view and the fact that many coherent emission mechanisms are brighter at lower frequencies.  SKA1-LOW would be the instrument of choice for FRB detection if their spectra are not flat below 1\,GHz and intergalactic scattering does not significantly broaden their pulse widths.

{\bf Incoherent search with SKA1-LOW}:  An initial survey is most plausibly conducted over the range extending from 200-350\,MHz, where the lowest frequency is limited to a relatively high value so that the effects of dispersion (and possibly scattering) are not so deleterious as to place onerous demands on computational capacity (and sensitivity).  If pulse scatter-broadening is large at these frequencies, a plausible alternative survey could be conducted over the frequency range 350-650\,MHz, if available.

Events can be detected by dedispersing and searching the co-added station beam powers.  (The sensitivity in 5\,ms for a 150 MHz bandwidth is 72 mJy.)  Upon detection of an event, this would trigger the automated dump of station voltages since the detected start time of the event; this mandates the presence of voltage buffering for a sizable fraction of each of SKA1-LOW's 1024 stations. A DM 2000 pc\,cm$^{-3}$ event would have a dispersion sweep lasting 140\,s for frequencies over the range 200-350 MHz. These data can either be processed online or offline to verify the event and determine its position; the volume of data to be dumped is not as large as might be supposed since one need extract only the subset of relevant frequency channels at the relevant times in which the signal will occur for a given dispersion measure.  Verification is made possible by the fact that a coherent summing of the station voltages at the time and dispersion measure of the event should yield a detection with far greater significance than that in the incoherent data stream. Since detection rate is linearly proportional to FoV, this survey approach would benefit enormously from the capability to form multiple independent station beams.  For instance, a transients detection system would capitalize on a system in which it were possible to trade beams for bandwidth.  Assuming a detection rate scaling as FoV $\times S_{\rm min}^{-3/2}$, one sees that a halving (an $n-th$ing) of the bandwidth but doubling of the FoV results in a net increase in the detection rate of $2^{1/4}$ ($n^{1/4}$).  A more moderate dependence on rate of FoV $\times S_{\rm min}^{-1}$ results in a detection rate increase of $2^{1/2}$ ($n^{1/2}$).

{\bf An alternative scenario is as above}, but instead the initial search is performed over a large ($\gtrsim 100$) number of tied-array beams formed from antennas in the dense central core (such as will be used for pulsar surveys).  This approach has the advantage that it would achieve a sensitivity $\sim 32$ times better than that of the incoherent search above at the possible expense of a smaller FoV.  This approach is superior to (1) if a sufficiently large number of tied-array beams are available that they cover more than 1/180th of the primary FoV\footnote{If the tied-array beams are formed from the core only, the necessary number of tied-array beams is $(2000/35)^2/180 = 18$.}, assuming a detection rate that scales as FoV$\times S_{\rm min}^{-3/2}$. 

{\bf A shallow all-sky search with SKA1-LOW}: An ultra-widefield search for transients would be capable of detecting events near the 1 Jy ms level at which they are currently detected by Parkes.  This mode would involve summing the powers of all 250000 antennas to search for events over the entire antenna FoV (roughly equivalent to 1/7th of the sky).  This yields an rms sensitivity of  800 mJy in 5 ms with 300 MHz bandwidth.  Coherent follow-up of these events could be performed by accessing the voltage data from just 1024 individual antennas.  Detection of an event triggers the capture of the voltage streams of 1024 single antennas, with an associated sensitivity of 400 mJy (5\,ms, 300 MHz) in order to perform event verification and localization.   However, calibration of the array in the presence of strong ionospheric fluctuations would present a challenge; one would need at least 256 antennas in the array to be closely-packed in order to perform ionospheric calibration, with the remainder providing the long baselines necessary to perform event localization.


\subsection{SKA1-SUR}
The attractiveness of SKA1-SUR is that it potentially permits searches over a range of different frequency bands, depending on which turns out to be optimal for FRB searches, though we note that at present only Band 2, covering 650-1670\,MHz, is expected to be deployed in Phase 1 of construction.
 
{\bf Incoherent search with SKA1-SUR}: a wide-field search for transients is effective with SKA1-SUR by searching the total powers for each PAF beam summed over all telescopes.  Upon detection of an event, the voltage buffers would be dumped and used to verify the event with a factor of $\sqrt{96} =9.7$ better S/N than the initial detection, and to localize it.  A buffer of duration $>$4\,s is required: for a DM=2000 pc cm$^{-3}$ event, the dispersion sweep across 300 MHz from 1.0 GHz to 1.3 GHz is 3.4s.  
We note that this is the approach advocated by the CRAFT survey consortium for detecting fast transients on ASKAP. 

{\bf Fast imaging search with SKA1-SUR}: the method above represents a relatively computationally light method of detecting FRBs that capitalizes on the full FoV of SKA1-SUR, but not its full sensitivity.  The ultimate capability would be provided by a fast (1ms-timescale) imaging mode, which would be capable of simultaneously detecting and localizing any FRBs.  However, we note that the requirement of searching each image pixel over a large range of trial dedispersions likely renders this approach computationally intractable at present.  This requirement is moderated by the fact that there is a factor 9.7 sensitivity advantage in an imaging survey relative to an incoherent survey.  An imaging survey would still be superior to an incoherent survey if one were able to integrate more coarsely in time and frequency with 50\,ms integration maps with 30$\times$10\,MHz channels.  The advantage of initially implementing a fast imaging mode is that one can improve resolution and thus sensitivity as computational power increases in future.

We note that such fast-imaging transient detection pipelines are already feasible and in use at present. 
One such pipeline runs on data from the VLA.  The code is parallelized to run in a multi-core, multi-node CPU environment simply by using tools available in Python. This system produces images at a rate of 20 images/s/core with a size of 1024$\times$1024 pixels (for 27 antennas, 2 polarizations and 256 MHz of bandwidth). This parallelizes well, so when running it on a cluster with 15 nodes of 15 cores, it sustains an imaging rate of 4500 images/s, enabling a survey which images 100\,DM trials with $\sim 25$\,ms integrations in real time.  Based on experience acquired at the VLA, this pipeline is feasible only for compact arrays; the current pipeline runs for VLA B-array configurations and smaller, namely 13000\,m$^2$ of collecting area within 10\,km.

\subsection{SKA1-MID}

The smaller instantaneous FoV of SKA1-MID compared to SKA1-LOW and SKA1-SUR puts it at a disadvantage in the detection of relatively rare and bright transient events.  Its nominal survey speed is at least a factor of two lower than that of SKA1-SUR.  However, this disadvantage can be partially mitigated if the backend hardware can exploit the full sensitivity of the array over its entire $\sim 1$sq.deg. field of view.  This may be possible if a pulsar backend proposed for SKA1-MID is constructed: this machine would be capable of forming $\approx 2200$ tied-array beams, covering the entire primary FoV of the antennas out to the half-power point of the beam.  This would give SKA1-MID an advantage over SKA1-SUR for more common, dimmer events.

\subsection{Relative detection rates of SKA1 variants}

It is instructive to compare the expected FRB detection rates of SKA1-LOW, SKA1-MID and SKA1-SUR.   In Table \ref{tab:compareSurvey} we compare the relative detection rates of various components of SKA1.  The absolute detection rate is obtained by noting that the Parkes detection rate is 1 FRB per $\approx 10\,$days, and that an incoherent survey mode with SKA1-SUR is expected to detect FRBs at a rate 6.5 times higher than Parkes, and a coherent (e.g. full sensitivity, full field of view) survey mode would detect them at a rate 197 times higher than Parkes.

\begin{table}[htbp!]
\begin{center}
\begin{tabular}{| c | ccc |}
\hline
Survey Metric & $\Omega S_0^{-1}$ & $\Omega S_0^{-3/2}$ & $\Omega S_0^{-2}$ \\ \hline
SKA1-SUR/MID (coherent) & 8.8 & 4.3 & 2.1 \\
SKA1-SUR/MID (incoherent) & 15 & 9.7 & 6.2 \\  \hline
SKA1-LOW/SKA1-SUR (coherent) & 3.8 & 6.3 & 10 \\
SKA1-LOW/SKA1-SUR (incoherent) & 1.2 & 1.1 & 0.9 \\ \hline
\end{tabular}
\end{center}
\caption{The relative detection rate of various components of SKA1 according to various possible scaling relations for the detection rate with sensitivity.  Here $S_0$ is the threshold detection flux density (for a given unspecified time interval) and $\Omega$ is the field of view.  In a Euclidean universe with no cosmological evolution of the FRB population the event rate scales as $\Omega S_0^{-3/2}$.  Flatter or steeper dependences on $S_0$ correspond to various plausible scenarios for the evolution of the FRB population throughout cosmic time.  The values computed here are derived on sensitivity and field of view figures tabled in the baseline design.  In making comparisons between various frequencies a flat spectral index is assumed and the possible effects of pulse smearing are neglected.  The contribution of $T_{\rm sky}$ to the system temperature below $\approx 200\,$MHz for SKA1-LOW is ignored, since FRBs are feasibly detected in the range 200-350\,MHz with this telescope.  A coherent survey mode corresponds to one in which one is capable of surveying the entire instrument primary FoV with the full array sensitivity, while the incoherent survey mode corresponds to one in which the total powers from each station/antenna are summed only.  It should be noted that this table forms only a part of the comparison between telescopes: the ability to implement a coherent detection mode is easier on telescopes such as SKA1-MID and SKA1-LOW relative to SKA1-SUR, and this makes SKA1-SUR less effective a transients detection instrument than first may be supposed.} \label{tab:compareSurvey}
\end{table}

It would seem from Table \ref{tab:compareSurvey} that SKA1-SUR is potentially inherently more effective at detecting FRBs at cm-wavelengths than SKA1-LOW, but that, if FRBs are detectable at low frequencies, a coherent survey mode on SKA1-LOW is at least half an order of magnitude more potent as a survey machine than SKA1-SUR.  

However, the comparison does not capture the ease of accessing the entire field of view at full sensitivity in the coherent survey mode; this is especially impractical for SKA1-SUR, with its low array filling factor, unless it is possible to implement a fast imaging mode.  Barring this eventuality, the most appropriate comparison at centimetre wavelengths is between an incoherent detection mode on SKA1-SUR and a coherent detection mode on SKA1-MID, for which a beamformer capable of tiling the approximately half the primary FoV is planned.  In this case the ratio of the expected detection rate on SKA1-MID relative to SKA1-SUR is either 0.9, 1.7 or 2.6, respectively for a detection rate that scales as either $S_0^{-1}$, $S_0^{-3/2}$ or $S_0^{-2}$.

In summary, SKA1-MID appears to be the immediate instrument of choice for FRB studies for three reasons: (1) it observes at the frequencies where FRBs have already been detected, (2) it has a compact core, so that its FoV can be utilized effectively, and (3) it has a planned pulsar backend capable of performing single-pulse searches.  

\subsection{The necessity of localization} \label{sec:localize}

In many instances, the scientific payoff from the detection of an astrophysical event is only realized once it is localized.  Event localization is paramount in the characterisation of one-off transient events, where the position is necessary to associate the event with a specific object or host galaxy.  This is necessary to determine its distance and thus energetics, and, through multi-wavelength follow-up, its origin.  The determination of burst positions has historically been central to the transients field, a key example being the use of accurate positions from the BeppoSAX satellite to resolve the thirty-year mystery of the origin of gamma-ray bursts.


Much of the transients science proposed for the SKA hinges upon its ability to localize events to sufficient accuracy to permit unambiguous association with a particular object.  In particular, the science case for FRBs hinges on the ability to localize them sufficiently well to identify their host galaxies and hence redshifts.  This requires event localization to within $\sim 0.1-0.5^{\prime \prime}$ for $z>1$ events.  Coeval photometric surveys such as DES and LSST will make host galaxy redshift determinations easy given their intention to identify most $z \leq 1.5$ galaxies over $\sim 50$\% of the sky.  We note that any detection today could be followed up with a 4m-class optical telescope to find a counterpart and get a redshift; the difficulty so far has been to obtain positions precise enough to merit optical follow-up.  

FRBs are but one class of fast transient object that require accurate localization.  The localization problem applies to every other fast transient that the SKA will detect.  For imaging surveys, applicable to slow transients the position is determined as part of the detection process.  However, this is not the case for fast transients, whose emission occurs below the integration timescale of the correlator.

Searching for short, highly dispersed signals in the imaging domain is computationally infeasible, as the image cubes require not only millisecond time resolution but also 1-100 kHz frequency resolution to avoid intra-channel smearing.  This implies $\sim 10^5$ images at every millisecond timestep.  The resultant $10^8$ images each second must then still be tested over many DM trials.
The most computationally feasible way to search for events on these timescales is to search station beam outputs or tied-array beams from the core of the array, which seek to maximize FoV and sensitivity, but not resolution.  Thus there is a fundamental disparity between the requirements of the transients search, which requires large FoV but poor resolution in order to be computationally tractable, and the event localization, which requires high resolution and only small FoV.

These computational realities necessitate the use of a transients buffer.  The role of the buffer is to store either station voltages from outer stations or high time-resolution correlation products on long baselines.  Once an event is detected in the low-resolution search, the corresponding information from the buffer, at the time of the event, is frozen.  A detailed search of the detection region is then performed at the specific epoch of the transient event in order to localize the event.

The use of a transients buffer fulfills a secondary role of event verification.  By including information from multiple distant stations, it is possible to verify that the event does not represent spurious emission from a local source of RFI.  

We emphasise that there will be many instances in which the localization of the emission from fast transients can be performed only in the radio domain.  This is because the emission from fast transients need not have an counterpart in any other part of the electromagnetic spectrum.  For emission of durations less than $\sim 1\,$s, the sensitivity of the SKA is such that for objects at $>$1kpc only objects with brightness temperatures well in excess of the inverse Compton limit for incoherent synchrotron emission are detectable.  Thus the emission from fast transients is necessarily coherent emission.  Coherent emission mechanisms operate most effectively at low frequencies, and usually do not extend above the radio wavelength regime.

These buffers would also add the important additional capability of enabling the telescope to follow-up possible triggers from transient events identified by telescopes in other wavebands.

\section{Early Science at Reduced Sensitivity} \label{sec:reducedSKA}

It is difficult to precisely assess the impact of reduced sensitivity on the FRB detection rate because the flux density distribution of FRBs is currently unknown.   However, several remarks are in order.
1.  The event rate likely scales in the range between $\Omega S_0^{-1}$ and $\Omega S_0^{-2}$.  Thus a 50\% reduction in the array sensitivity will cause a decrease in the event detection rate between a factor of two and a factor of four.  Even with this reduction, the telescope will be much more sensitive than Parkes, so there will be no problem detecting FRBs.  2.  We still require the commensal transients capability in this early phase and the ability to localize the events.  

\section{Science for the Full SKA} \label{sec:fullSKA}

There is currently so much important information lacking about the nature of FRBs that it is difficult to assess the impact of the transition from SKA1 to the full SKA, other than to say that the increased sensitivity, coherent field-of-view and processing power will open up for investigation huge new volumes of parameter space; new discoveries are almost bound to occur. Early results from SKA-precursors \citep[e.g.][]{Trott13} and SKA1 will determine the source count and spectrum of FRBs and so will allow critical design decisions for the full SKA to reflect FRB search requirements. One such driver to be assessed will be the importance of the intrinsic hemispherical sensitivity offered by the Mid-Frequency Aperture Array for identifying fast transient events in the Universe.   



\section{Conclusions} \label{sec:conclusions}

The occurrence of impulsive radio transients at cosmological distances, as embodied by FRBs, offers enormous scope for addressing a number of cosmological questions of fundamental significance.  This includes direction detection and characterisation of the ``missing'' baryons in the Inter-Galactic Medium.  The prospect of probing the dark energy equation of state will be possible if these events can be detected at redshifts $z \gtrsim 2$.  
 
The ability to perform this science is contingent upon several facets of the design of the SKA.  The telescope must be able to detect events in sufficient numbers (i.e. $>10^3$), and to do this it must be able to access a large field of view for long periods of time (i.e.\,as part of a commensal survey).   It must be able to localize events with sufficient accuracy to enable the measurement of redshifts at optical wavelengths; this facility is most easily provided by transient buffers, which can localize an event after each burst is detected.  This also entails the provision of the ability to trigger the localization through the telescope signal path.  

Finally, we remark that the relative infancy of this field means that it is likely that many more significant discoveries await, and it is not possible to specify precisely telescope all the requirements that best suit this future science.  However, by adopting a modular design approach which ensures access to key points in the telescope signal path are not designed out will greatly increase the scope to exploit future discoveries as they are made over the fifty year lifetime of the telescope.

\setlength{\bibsep}{0.0pt}
\bibliographystyle{apj}

\end{document}